\documentclass[prd,showpacs,showkeys,amsmath,amssymb]{revtex4}
\usepackage{hyperref,slashed,graphicx,color}
\begin{document}
\title{Derivation of Electroweak Chiral Lagrangian
from One Family Technicolor Model}
\author{Hong-Hao Zhang$^{1}$}
\author{Kai-Xi Feng$^{1}$}
\author{Shao-Zhou Jiang$^2$}
\author{Qing Wang$^2$}
\affiliation{%
$^1$School of Physics $\&$ Engineering, Sun Yat-Sen University,
Guangzhou
510275, China\\
$^2$Department of Physics $\&$ Center for High Energy Physics,
Tsinghua University, Beijing 100084, China}

\begin{abstract}
Based on previous studies deriving the chiral Lagrangian for pseudo
scalar mesons from the first principle of QCD in the path integral
formalism, we derive the electroweak chiral Lagrangian and
dynamically compute all its coefficients from the one family
technicolor model. The numerical results of the $p^4$ order
coefficients obtained in this paper are proportional to the
technicolor number $N_{\rm TC}$ and the technifermion number $N_{\rm
TF}$, which agrees with the arguments in previous works, and which
confirms the reliability of this dynamical computation.
\end{abstract}
 \maketitle
%%%%%%%%%%%%%%%%%%%%%%%%%%%%%%%%%%%%%%%%%%%%%%%%%%%%%%

So far the postulated Higgs particle of the standard model was not
observed in experiments. We do not know what the electroweak
symmetry breaking mechanism in nature is. The electroweak chiral
Lagrangian (EWCL) is a general low-energy description for
electroweak symmetry breaking patterns
\cite{Appelquist:1980vg,Appelquist:1993ka}, especially for those
strongly breaking electroweak symmetry mechanisms. All the
coefficients of the EWCL, in principle, can be fixed by experiments.
Due to the nonperturbative property of the possible strong dynamics,
it is difficult to relate the measured coefficients of the EWCL to
underlying theories, except for some special coefficients like the
$S$ parameter which may be estimated from dispersion relations
\cite{Peskin:1990zt,Peskin:1991sw}. Recently, the series of work of
Ref.
\cite{Wang:1999cp,Wang:1999xh,Wang:2000mg,Wang:2000mu,Yang:2002he,
Wang:2002rb,Wang:2005cm,Ma:2003uv} successfully produced the
predictions for the coefficients of the chiral Lagrangian from the
underlying theory of QCD, which lights up the hope of building up
the relationship between the coefficients of the EWCL and underlying
strong dynamical models. Along this line of thought, we have built
up a new dynamical computation formalism and derived the EWCL and
dynamically compute all its coefficients, respectively, from the one
doublet and topcolor-assisted technicolor models
\cite{Zhang:2007zi}. As a nontrivial application of this dynamical
computation formalism, in this paper we will derive the EWCL and
compute its coefficients from the one family technicolor model.

Let us consider the one-family technicolor model
\cite{Farhi:1980xs,Hill:2002ap}. The global symmetry breaking
pattern of this model is $SU(8)\times SU(8)/SU(8)$. Besides the 3
Goldstone bosons eaten by the electroweak gauge bosons, there are
extra 60 Goldstone bosons. Since those 60 Goldstone bosons were not
observed in experiments, they might either be eaten by more massive
gauge bosons, or be massive as pseudo-Goldstone bosons due to some
explicit symmetry breaking terms. For simplicity and as the first
step of the probe, in the following we only consider the effects of
the 3 Goldstone bosons eaten by the electroweak gauge bosons. Thus,
the unitary unimodular matrix $U$-field in the electroweak chiral
Lagrangian is still $2\times 2$ and spanned by these 3 would-be
Goldstone bosons. In this model, the technifermions are assigned  to
the gauge group $G_{\rm TC}\times SU(3)_C\times SU(2)_L\times
U(1)_Y$ as follows:
\begin{eqnarray}
&&Q_L=\begin{pmatrix}U\\
D\end{pmatrix}_L\sim(R,~3,~2,~\frac{1}{6})\;,\nonumber\\
&& U_R\sim(R,~3,~1,~\frac{2}{3})\;,\qquad
D_R\sim(R,~3,~1,~-\frac{1}{3})\;,\nonumber\\
&&L_L=\begin{pmatrix}N\\
E\end{pmatrix}_L\sim(R,~1,~2,~-\frac{1}{2})\;,\nonumber\\
&& N_R\sim(R,~1,~1,~0)\;,\qquad E_R\sim(R,~1,~1,~-1)\;.
\end{eqnarray}
With these assignments, the chiral techniquarks $Q_{L,R}$ and the
technileptons $L_{L,R}$ are assumed to transform according to the
same complex representation $R$ of $G_{\rm TC}$, and they have the
same electric charges defined by $Q=T_3+Y$ as ordinary quarks and
leptons. Neglecting the contributions of ordinary fermions, we focus
on the action of technifermions, technigluons, gluons and
electroweak-gauge-bosons' sector, {\it i.e.} the electroweak
symmetry breaking sector (SBS) of this model,
\begin{eqnarray}
S_{\rm SBS}&=&\int d^4x\bigg[-\frac{1}{4}F_{\mu\nu}^\alpha
F^{\alpha,\mu\nu}-\frac{1}{4}A_{\mu\nu}^A
A^{A,\mu\nu}-\frac{1}{4}W_{\mu\nu}^aW^{a,\mu\nu}
-\frac{1}{4}B_{\mu\nu}B^{\mu\nu}\nonumber\\
&&+\bar{Q}\bigg(i\slashed{\partial}-g_{\rm TC}t^\alpha
\slashed{G}^\alpha-g_3\frac{\lambda^A}{2}\slashed{A}^A-g_2\frac{\tau^a}{2}\slashed{W}^a
P_L-g_1(\frac{\tau^3}{2}P_R+\frac{1}{6})\slashed{B}\bigg)Q\nonumber\\
&&+\bar{L}\bigg(i\slashed{\partial}-g_{\rm TC}t^\alpha
\slashed{G}^\alpha-g_2\frac{\tau^a}{2}\slashed{W}^a
P_L-g_1(\frac{\tau^3}{2}P_R-\frac{1}{2})\slashed{B}\bigg)L\bigg]\,.
\end{eqnarray}
where $g_{\rm TC}$, $g_3$, $g_2$ and $g_1$ ($G_\mu^\alpha$,
$A_\mu^A$, $W_\mu^a$ and $B_\mu$) are the coupling constants (gauge
fields) of $G_{\rm TC}\times SU(3)_C\times SU(2)_L\times U(1)_Y$
with technicolor index $\alpha$ ($\alpha=1,2,\ldots,N^2-1$), color
index $A$ ($A=1,2,\ldots,8$) and weak index $a$ ($a=1,2,3$)
respectively; and $F_{\mu\nu}^\alpha$, $A_{\mu\nu}^A$,
$W_{\mu\nu}^a$ and $B_{\mu\nu}$ are the corresponding field strength
tensors; $t^\alpha$ ($\alpha=1,2,\ldots,N^2-1$) are the generators
for the representation $R$ of $G_{\rm TC}$, while $\lambda^A$
($A=1,2,\ldots,8$) and $\tau^a$ ($a=1,2,3$) are respectively
Gell-Man and Pauli matrices; and the left and right chirality
projection operators are $P{\substack{\tiny
L\\R}}\equiv(1\mp\gamma_5)/2$. To derive the low-energy effective
Lagrangian -- the EWCL -- from this fundamental technicolor model,
we need to integrate out technigluons and technifermions above the
weak scale which can be formulated as
\begin{eqnarray}
\int{\cal D}G_\mu^\alpha{\cal D}A_\mu^A{\cal D}\bar{\psi}{\cal
D}\psi\exp\bigg(\,iS_{\rm
SBS}\big[G_\mu^\alpha,A_\mu^A,W_\mu^a,B_\mu,\bar{\psi},\psi\big]\bigg)
=\int{\cal D}\mu(U)\exp\bigg(iS_{\rm
eff}[U,W_\mu^a,B_\mu]\bigg)\;,\label{strategy}
\end{eqnarray}
where $\psi$ denotes the technifermions including $Q$ and $L$,
$U(x)$ is a dimensionless unitary unimodular matrix field in the
EWCL, and ${\cal D}\mu(U)$ denotes the normalized functional
integration measure. After integrating out technigluons and gluons
by means of the functional integral approach similar to Ref.
\cite{Zhang:2007zi}, we can obtain $S_{\rm eff}[U,W_\mu^a,B_\mu]$
divided into gauge field kinetic-energy sector, normal and anomalous
parts:
\begin{eqnarray}
S_{\rm eff}[U,W_\mu^a,B_\mu]= \int
d^4x(-\frac{1}{4}W_{\mu\nu}^aW^{a,\mu\nu}-\frac{1}{4}B_{\mu\nu}B^{\mu\nu})
+S_{\rm norm}[U,W_\mu^a,B_\mu]+S_{\rm anom}[U,W_\mu^a,B_\mu]\;,
\end{eqnarray}
where the normal part of the effective action, $S_{\rm
norm}[U,W_\mu^a,B_\mu]$, may be written as
\begin{eqnarray}
&&\hspace{-0.5cm}\exp\bigg(iS_{\rm
norm}[U,W_\mu^a,B_\mu]\bigg)\nonumber\\
&=&\int\mathcal{D}\bar{Q}_\xi\mathcal{D}Q_\xi
\mathcal{D}\bar{L}_\xi\mathcal{D}L_\xi\exp\bigg\{i\int
d^4x\bigg[\bar{Q}_\xi\bigg(i\slashed{\partial}-g_2\frac{\tau^a}{2}\slashed{W}_\xi^a
P_L-g_1(\frac{\tau^3}{2}P_R+\frac{1}{6})\slashed{B}_\xi
-\Sigma(\overline{\nabla}^2)\bigg)Q_\xi\nonumber\\
&&+\bar{L}_\xi\bigg(i\slashed{\partial}-g_2\frac{\tau^a}{2}\slashed{W}_\xi^a
P_L-g_1(\frac{\tau^3}{2}P_R-\frac{1}{2})\slashed{B}_\xi
-\Sigma(\overline{\nabla}^2)\bigg)L_\xi\bigg]\bigg\}\nonumber\\
&=&\int\mathcal{D}\bar{Q}_\xi\mathcal{D}Q_\xi
\mathcal{D}\bar{L}_\xi\mathcal{D}L_\xi\exp\bigg\{i\int
d^4x\bigg[\bar{Q}_\xi\bigg(i\slashed{\partial}
+\slashed{\tilde{v}}+\slashed{\tilde{a}}\gamma_5
-\Sigma(\overline{\nabla}^2)\bigg)Q_\xi
+\bar{L}_\xi\bigg(i\slashed{\partial}+\slashed{\tilde{v}}^\prime
+\slashed{\tilde{a}}\gamma_5
-\Sigma(\overline{\nabla}^2)\bigg)L_\xi\bigg]\bigg\}\nonumber\\
&=&\int\mathcal{D}\bar{\psi}_\xi\mathcal{D}\psi_\xi\exp\bigg[i\int
d^4x\bar{\psi}_\xi\bigg(i\slashed{\partial}+\slashed{\tilde{V}}+\slashed{\tilde{A}}\gamma_5
-\Sigma(\overline{\nabla}^2)\bigg)\psi_\xi\bigg]\;,
\end{eqnarray}
with
\begin{eqnarray}
&&\tilde{v}_\mu\equiv-\frac{1}{2}g_2\frac{\tau^a}{2}W_{\xi,\mu}^a
-\frac{1}{2}g_1\frac{\tau^3}{2}B_{\xi,\mu}-\frac{1}{6}g_1B_{\xi,\mu}\;,\nonumber\\
&&\tilde{a}_\mu\equiv\frac{1}{2}g_2\frac{\tau^a}{2}W_{\xi,\mu}^a
-\frac{1}{2}g_1\frac{\tau^3}{2}B_{\xi,\mu}\;,\nonumber\\
&&\tilde{v}_\mu^\prime\equiv-\frac{1}{2}g_2\frac{\tau^a}{2}W_{\xi,\mu}^a
-\frac{1}{2}g_1\frac{\tau^3}{2}B_{\xi,\mu}+\frac{1}{2}g_1B_{\xi,\mu}\;,\nonumber\\
&&\psi_\xi\equiv\begin{pmatrix}Q_\xi^r\\Q_\xi^g\\Q_\xi^b\\L_\xi\end{pmatrix}\;,\qquad
\tilde{V}_\mu\equiv\begin{pmatrix}\tilde{v}_\mu & & &\\ &
\tilde{v}_\mu & & \\ & & \tilde{v}_\mu & \\ & & &
\tilde{v}_\mu^\prime\end{pmatrix}\;,\nonumber\\
&&\tilde{A}_\mu\equiv\begin{pmatrix}\tilde{a}_\mu & & &\\ &
\tilde{a}_\mu & & \\ & & \tilde{a}_\mu & \\ & & &
\tilde{a}_\mu\end{pmatrix}\;,
\end{eqnarray}
or
\begin{eqnarray}
&&\tilde{V}_\mu=(-\frac{1}{2}g_2\frac{\tau^a}{2}W_{\xi,\mu}^a
-\frac{1}{2}g_1\frac{\tau^3}{2}B_{\xi,\mu})\mathbf{I}
-\frac{1}{2}g_1B_{\xi,\mu}\mathbf{I_F}\;,\nonumber\\
&&\tilde{A}_\mu=(\frac{1}{2}g_2\frac{\tau^a}{2}W_{\xi,\mu}^a
-\frac{1}{2}g_1\frac{\tau^3}{2}B_{\xi,\mu})\mathbf{I}\;,\label{Adef-1f}
\end{eqnarray}
\begin{eqnarray}
\mathbf{I}=\begin{pmatrix}1 & & &\\ & 1 & &
\\ & & 1 & \\ & & & 1\end{pmatrix}\;,\qquad
\mathbf{I_F}=\begin{pmatrix}\frac{1}{3} & & &\\ & \frac{1}{3} & &
\\ & & \frac{1}{3} & \\ & & & -1\end{pmatrix}\;.
\end{eqnarray}
Here we have made the chiral decomposition of $U(x)$ as
$U(x)\equiv\xi_L^\dag(x)\xi_R(x)$, with the unitary matrices
$\xi_L(x)\equiv e^{i\tau^a\phi^a}$ and $\xi_R(x)\equiv
e^{i\tau^3\phi^0}$ respectively for the $SU(2)_L$ and the $U(1)_Y$
phase degrees of freedom. And the chiral rotated technifermions and
gauge fields are respectively defined by
\begin{eqnarray}
&&\psi_\xi(x)=P_L\xi_L(x)\psi_L(x)+P_R\xi_R(x)\psi_R(x)\,,
\nonumber\\
&&g_2\frac{\tau^a}{2}W_{\xi,\mu}^a(x)\equiv
\xi_L(x)[g_2\frac{\tau^a}{2}W_{\mu}^a(x)-i\partial_\mu]\xi_L^\dag(x)\,,\nonumber\\
&&g_1\frac{\tau^3}{2}B_{\xi,\mu}(x)\equiv
\xi_R(x)[g_1\frac{\tau^3}{2}B_{\mu}(x)-i\partial_\mu]\xi_R^\dag(x)\,.
\end{eqnarray}
Now we can parameterize $S_{\rm norm}[U,W_\mu^a,B_\mu]$ as follows:
\begin{eqnarray}
&&\hspace{-0.5cm}iS_{\rm
norm}[U,W_\mu^a,B_\mu]=\mathrm{Tr}\log[i\slashed{\partial}+\slashed{\tilde{V}}
+\slashed{\tilde{A}}\gamma_5-\Sigma(\overline{\nabla}^2)]\nonumber\\
&=&i\int d^4x\,\mathrm{tr}_f\bigg[(F_0^{\rm
1F})^2\tilde{A}^2-\mathcal{K}_1^{\rm 1F}(d_\mu
\tilde{A}^\mu)^2-\mathcal{K}_2^{\rm 1F}(d_\mu \tilde{A}_\nu-d_\nu
\tilde{A}_\mu)^2+\mathcal{K}_3^{\rm
1F}(\tilde{A}^2)^2+\mathcal{K}_4^{\rm
1F}(\tilde{A}_\mu\tilde{A}_\nu)^2-\mathcal{K}_{13}^{\rm
1F}\tilde{V}_{\mu\nu}\tilde{V}^{\mu\nu} \nonumber\\
&&+i\mathcal{K}_{14}^{\rm 1F}\tilde{A}_\mu \tilde{A}_\nu
\tilde{V}^{\mu\nu}\bigg]+\mathcal{O}(p^6)\;,\label{action-GND-1fam-norm}
\end{eqnarray}
where $d_\mu \tilde{A}_\nu\equiv\partial_\mu
\tilde{A}_\nu-i[\tilde{V}_\mu, \tilde{A}_\nu]$,
$\tilde{V}_{\mu\nu}\equiv i[\partial_\mu-i\tilde{V}_\mu,
\partial_\nu-i\tilde{V}_\nu]$, and the parameters $\mathcal{K}_i^{\rm
1F}$ (where the superscript 1F stands for the one family technicolor
model) are functions of the technifermion self-energy $\Sigma(p^2)$
determined by the Schwinger-Dyson equations. Their detailed
expressions have already been given in Eq. (36) of Ref.
\cite{Yang:2002he} in the QCD case just with the replacement of
$N_c\rightarrow {\rm rank}(G_{\rm TC})$ in the present case. On the
other hand, the anomalous part $S_{\rm anom}[U,W_\mu^a,B_\mu]$ may
also be parameterized as
\begin{eqnarray}
&&\hspace{-0.5cm}iS_{\rm
anom}[U,W_\mu^a,B_\mu]=\mathrm{Tr}\log[i\slashed{\partial}+\slashed{\tilde{V}}
+\slashed{\tilde{A}}\gamma_5]\bigg|_{\xi_L=\xi_R=1}-\mathrm{Tr}\log[i\slashed{\partial}
+\slashed{\tilde{V}}
+\slashed{\tilde{A}}\gamma_5]\nonumber\\
&=&\mathrm{Tr}\log[i\slashed{\partial}+\slashed{\tilde{V}}
+\slashed{\tilde{A}}\gamma_5]\bigg|_{\xi_L=\xi_R=1} +i\int
d^4x\,\mathrm{tr}_f\bigg[-\mathcal{K}_1^{\rm 1F,(anom)}(d_\mu
\tilde{A}^\mu)^2-\mathcal{K}_2^{\rm 1F,(anom)}(d_\mu
\tilde{A}_\nu-d_\nu
\tilde{A}_\mu)^2 \nonumber\\
&&+\mathcal{K}_3^{\rm 1F,(anom)}(\tilde{A}^2)^2+\mathcal{K}_4^{\rm
1F,(anom)}(\tilde{A}_\mu\tilde{A}_\nu)^2-\mathcal{K}_{13}^{\rm
1F,(anom)}\tilde{V}_{\mu\nu}\tilde{V}^{\mu\nu}+i\mathcal{K}_{14}^{\rm
1F,(anom)}\tilde{A}_\mu \tilde{A}_\nu
\tilde{V}^{\mu\nu}\bigg]+\mathcal{O}(p^6)\;,
\end{eqnarray}
with the parameters ${\cal K}_i^{\rm 1F,(anom)}=-\big[{\cal
K}_i^{\rm 1F}\big]_{\Sigma=0}$ ($i=1,2,3,4,13,14$). Finally, we
obtain
\begin{eqnarray}
&&\hspace{-0.5cm}S_{\rm eff}[U,W_\mu^a,B_\mu]=-\frac{1}{4}\int
d^4x(W_{\mu\nu}^aW^{a,\mu\nu}+B_{\mu\nu}B^{\mu\nu})+\mathrm{Tr}\log[i\slashed{\partial}
+\slashed{\tilde{V}}
+\slashed{\tilde{A}}\gamma_5]\bigg|_{\xi_L=\xi_R=1}\nonumber\\
&&+ i\int d^4x\,\mathrm{tr}_f\bigg[(F_0^{\rm
1F})^2\tilde{A}^2-\mathcal{K}_1^{{\rm 1F},\Sigma\neq0}(d_\mu
\tilde{A}^\mu)^2-\mathcal{K}_2^{{\rm 1F},\Sigma\neq0}(d_\mu
\tilde{A}_\nu-d_\nu \tilde{A}_\mu)^2+\mathcal{K}_3^{{\rm
1F},\Sigma\neq0}(\tilde{A}^2)^2 \nonumber\\
&&+\mathcal{K}_4^{{\rm
1F},\Sigma\neq0}(\tilde{A}_\mu\tilde{A}_\nu)^2-\mathcal{K}_{13}^{{\rm
1F},\Sigma\neq0}\tilde{V}_{\mu\nu}\tilde{V}^{\mu\nu}+i\mathcal{K}_{14}^{{\rm
1F},\Sigma\neq0}\tilde{A}_\mu \tilde{A}_\nu
\tilde{V}^{\mu\nu}\bigg]+\mathcal{O}(p^6)\;,\label{action-GND-1fam-full}
\end{eqnarray}
where $\mathcal{K}_i^{{\rm 1F},\Sigma\neq0}$ are the
$\Sigma$-dependent parts of $\mathcal{K}_i$, that is,
$\mathcal{K}_i^{{\rm 1F},\Sigma\neq0}\equiv\mathcal{K}_i^{\rm
1F}-\big[\mathcal{K}_i^{\rm 1F}\big]_{\Sigma=0}$
($i=1,2,3,4,13,14$). To build the explicit relationship between
terms of Eq. (\ref{action-GND-1fam-full}) and terms of the EWCL, we
need to derive some identities on the chiral rotated gauge fields.
First of all, from Eq. (\ref{Adef-1f}), we obtain
\begin{eqnarray}
\tilde{A}_\mu=-\frac{i}{2}\xi_R
X_\mu\xi_R^\dag\,\mathbf{I}\label{appendix-rotated-A-til-mu}
\end{eqnarray}
where $X_\mu\equiv U^\dag(D_\mu U)$, and
\begin{eqnarray}
\tilde{V}_\mu=\frac{i}{2}\xi_RX_\mu\xi_R^\dag\mathbf{I}
-\xi_R(g_1\frac{\tau^3}{2}B_\mu-i\partial_\mu)\xi_R^\dag\mathbf{I}
-\frac{1}{2}[g_1B_\mu-i\mathrm{tr}(\tau^3\xi_R\partial_\mu\xi_R^\dag)]\mathbf{I_F}\;,
\nonumber
\end{eqnarray}
which leads to the following two relations:
\begin{eqnarray}
i\xi_R^\dag\tilde{V}_\mu&=&-\frac{1}{2}X_\mu\xi_R^\dag\mathbf{I}
-ig_1\frac{\tau^3}{2}B_\mu\xi_R^\dag\mathbf{I}-(\partial_\mu\xi_R^\dag)\mathbf{I}
-\frac{i}{2}[g_1B_\mu-i\mathrm{tr}(\tau^3\xi_R\partial_\mu\xi_R^\dag)]
\xi_R^\dag\mathbf{I_F}\;,\nonumber\\
-i\tilde{V}_\mu\xi_R&=&\frac{1}{2}\xi_RX_\mu\mathbf{I}
+i\xi_Rg_1\frac{\tau^3}{2}B_\mu\mathbf{I}-(\partial_\mu\xi_R)\mathbf{I}
+\frac{i}{2}[g_1B_\mu-i\mathrm{tr}(\tau^3\xi_R\partial_\mu\xi_R^\dag)]
\xi_R\mathbf{I_F}\;,\nonumber
\end{eqnarray}
{\it i.e.},
\begin{eqnarray}
&&(\partial_\mu\xi_R^\dag)\mathbf{I}+i\xi_R^\dag\tilde{V}_\mu=
(-\frac{1}{2}X_\mu-ig_1\frac{\tau^3}{2}B_\mu)\xi_R^\dag\mathbf{I}
-\frac{i}{2}[g_1B_\mu-i\mathrm{tr}(\tau^3\xi_R\partial_\mu\xi_R^\dag)]
\xi_R^\dag\mathbf{I_F}\;,\nonumber\\
&&(\partial_\mu\xi_R)\mathbf{I}-i\tilde{V}_\mu\xi_R=
\xi_R(\frac{1}{2}X_\mu+ig_1\frac{\tau^3}{2}B_\mu)\mathbf{I}
+\frac{i}{2}[g_1B_\mu-i\mathrm{tr}(\tau^3\xi_R\partial_\mu\xi_R^\dag)]
\xi_R\mathbf{I_F}\;.\nonumber
\end{eqnarray}
Thus, for any chiral rotated field $f\equiv\xi_R F\xi_R^\dag$, we
have
\begin{eqnarray}
d_\mu f&\equiv&\partial_\mu f-i[\tilde{V}_\mu, f]=\xi_R(\partial_\mu
F)\xi_R^\dag+[(\partial_\mu\xi_R)\mathbf{I}-i\tilde{V}_\mu\xi_R]F\xi_R^\dag
+\xi_RF[(\partial_\mu\xi_R^\dag)\mathbf{I}+i\xi_R^\dag\tilde{V}_\mu]\nonumber\\
&=&\xi_R\bigg\{(D_\mu F)+\frac{1}{2}[X_\mu, F]\bigg\}\xi_R^\dag
+\frac{i}{2}[g_1B_\mu-i\mathrm{tr}(\tau^3\xi_R\partial_\mu\xi_R^\dag)]
\xi_R[\mathbf{I_F}, F]\xi_R^\dag\;.
\end{eqnarray}
In particular, if $f=\tilde{A}_\nu=-\frac{i}{2}\xi_R
X_\nu\xi_R^\dag\,\mathbf{I}$, then $F=-\frac{i}{2}X_\nu\mathbf{I}$
and we have
\begin{eqnarray}
d_\mu\tilde{A}_\nu=-\frac{i}{2}\xi_R\big\{(D_\mu
X_\nu)+\frac{1}{2}[X_\mu,X_\nu]\big\}\xi_R^\dag\,\mathbf{I}
\label{appendix-rotated-d-mu-A-nu-1fam}
\end{eqnarray}
which implies
\begin{eqnarray}
&&d_\mu\tilde{A}_\nu-d_\nu\tilde{A}_\mu=
\frac{1}{2}\xi_R\big(\overline{W}_{\mu\nu}-g_1\frac{\tau^3}{2}B_{\mu\nu}
\big)\xi_R^\dag\,\mathbf{I}\;,\\
&&\mathrm{tr}[(d_\mu\tilde{A}_\nu-d_\nu\tilde{A}_\mu)^2]=
\mathrm{tr}(\overline{W}_{\mu\nu}\overline{W}^{\mu\nu})
-g_1B_{\mu\nu}\mathrm{tr}(\tau^3\overline{W}^{\mu\nu})
+\frac{1}{2}g_1^2B_{\mu\nu}B^{\mu\nu}\;.
\end{eqnarray}
To proceed, let us show an identity on chiral rotated gauge field
strength:
\begin{eqnarray}
\tilde{V}_{\mu\nu}=\xi_R\bigg[\bigg(-\frac{1}{2}\overline{W}_{\mu\nu}
-\frac{1}{2}g_1\frac{\tau^3}{2}B_{\mu\nu}
-\frac{i}{4}[X_\mu,X_\nu]\bigg)\mathbf{I}
-\frac{1}{2}g_1B_{\mu\nu}\mathbf{I_F}\bigg]\xi_R^\dag
\label{appendix-rotated-V-mu-nu-1fam}
\end{eqnarray}
{\bf Proof\,:} To compute the rotated gauge field strength
$\tilde{V}_{\mu\nu}$, let us decompose $\tilde{V}_\mu$ as
$\tilde{V}_\mu=\tilde{V}_\mu^\prime+\tilde{S}_\mu$, where
\begin{eqnarray*}
\tilde{V}_\mu^\prime\equiv\xi_R\bigg[(\frac{i}{2}X_\mu-g_1\frac{\tau^3}{2}B_\mu)\mathbf{I}
-\frac{1}{2}g_1B_\mu\mathbf{I_F}\bigg]\xi_R^\dag\;,\qquad
\tilde{S}_\mu\equiv\xi_R(\partial_\mu\xi_R^\dag)\mathbf{I}
+\frac{i}{2}\mathrm{tr}(\tau^3\xi_R\partial_\mu\xi_R^\dag)\mathbf{I_F}\;.
\end{eqnarray*}
It is easily checked that
$\partial_\mu\tilde{S}_\nu-\partial_\nu\tilde{S}_\mu=0$ and
$[\tilde{S}_\mu,\tilde{S}_\nu]=0$. Thus,
\begin{eqnarray*}
\tilde{V}_{\mu\nu}=d_\mu\tilde{V}_\nu^\prime-d_\nu\tilde{V}_\mu^\prime
+i[\tilde{V}_\mu^\prime,\tilde{V}_\nu^\prime]=\xi_R\bigg[\bigg(-\frac{1}{2}\overline{W}_{\mu\nu}
-\frac{1}{2}g_1\frac{\tau^3}{2}B_{\mu\nu}
-\frac{i}{4}[X_\mu,X_\nu]\bigg)\mathbf{I}
-\frac{1}{2}g_1B_{\mu\nu}\mathbf{I_F}\bigg]\xi_R^\dag\;.\qquad \Box
\end{eqnarray*}
Eq. \eqref{appendix-rotated-V-mu-nu-1fam} gives
\begin{eqnarray}
\mathrm{tr}(\tilde{V}_{\mu\nu}\tilde{V}^{\mu\nu})
&=&\mathrm{tr}(\overline{W}_{\mu\nu}\overline{W}^{\mu\nu})
+\frac{7}{6}g_1^2B_{\mu\nu}B^{\mu\nu}-\frac{1}{2}[\mathrm{tr}(X_\mu
X_\nu)]^2+\frac{1}{2}[\mathrm{tr}(X_\mu X^\mu)]^2
+g_1B_{\mu\nu}\mathrm{tr}(\tau^3\overline{W}^{\mu\nu})\nonumber\\
&&+2i\mathrm{tr}(X_\mu X_\nu\overline{W}^{\mu\nu})
+ig_1B_{\mu\nu}\mathrm{tr}(\tau^3X^\mu X^\nu)\;,
\end{eqnarray}
and
\begin{eqnarray}
\mathrm{tr}(\tilde{A}_{\mu}\tilde{A}_{\nu}\tilde{V}^{\mu\nu})
&=&\frac{1}{2}\mathrm{tr}(X_\mu X_\nu\overline{W}^{\mu\nu})
+\frac{1}{4}g_1B_{\mu\nu}\mathrm{tr}(\tau^3X^\mu
X^\nu)+\frac{i}{4}[\mathrm{tr}(X_\mu
X_\nu)]^2\nonumber\\
&&-\frac{i}{4}[\mathrm{tr}(X_\mu X^\mu)]^2\;.\label{rotated-last}
\end{eqnarray}
Now, using the relations
(\ref{appendix-rotated-A-til-mu}--\ref{rotated-last}), we can
reexpress Eq.\eqref{action-GND-1fam-full} term by term as follows:
\begin{eqnarray}
&&(F_0^{\rm 1F})^2\mathrm{tr}_f(\tilde{A}^2)=-(F_0^{\rm
1F})^2\mathrm{tr}_f(X_\mu X^\mu)\;,\\
&&\mathcal{K}_3^{{\rm
1F},\Sigma\neq0}\mathrm{tr}_f[(\tilde{A}^2)^2]=\frac{\mathcal{K}_3^{{\rm
1F},\Sigma\neq0}}{4}\mathrm{tr}_f(X_\mu X^\mu X_\nu
X^\nu)=\frac{\mathcal{K}_3^{{\rm
1F},\Sigma\neq0}}{8}[\mathrm{tr}_f(X_\mu X^\mu)]^2\;,\\
&&\mathcal{K}_4^{{\rm 1F},\Sigma\neq0}\mathrm{tr}_f[(\tilde{A}_\mu
\tilde{A}_\nu)^2]=\frac{\mathcal{K}_4^{{\rm
1F},\Sigma\neq0}}{4}\mathrm{tr}_f(X_\mu X_\nu X^\mu
X^\nu)=\frac{\mathcal{K}_4^{{\rm
1F},\Sigma\neq0}}{4}[\mathrm{tr}_f(X_\mu
X_\nu)]^2-\frac{\mathcal{K}_4^{{\rm
1F},\Sigma\neq0}}{8}[\mathrm{tr}_f(X_\mu X^\mu)]^2\;,\\
&&d_\mu \tilde{A}^\mu=-\frac{i}{2}\xi_R\big\{(D_\mu
X^\mu)+\frac{1}{2}[X_\mu,X^\mu]\big\}\xi_R^\dag\,\mathbf{I}=0\;,\\
&&-\mathcal{K}_2^{{\rm 1F},\Sigma\neq0}\mathrm{tr}_f[(d_\mu
\tilde{A}_\nu-d_\nu \tilde{A}_\mu)^2]=-\mathcal{K}_2^{{\rm
1F},\Sigma\neq0}\big[\mathrm{tr}_f(\overline{W}_{\mu\nu}\overline{W}^{\mu\nu})
-g_1B_{\mu\nu}\mathrm{tr}_f(\tau^3\overline{W}^{\mu\nu})
+\frac{1}{2}g_1^2B_{\mu\nu}B^{\mu\nu}\big]\;,\\
&&-\mathcal{K}_{13}^{{\rm
1F},\Sigma\neq0}\mathrm{tr}_f(\tilde{V}_{\mu\nu}\tilde{V}^{\mu\nu})
=-\mathcal{K}_{13}^{{\rm 1F},\Sigma\neq0}\bigg[
\mathrm{tr}(\overline{W}_{\mu\nu}\overline{W}^{\mu\nu})
+\frac{7}{6}g_1^2B_{\mu\nu}B^{\mu\nu}-\frac{1}{2}[\mathrm{tr}(X_\mu
X_\nu)]^2+\frac{1}{2}[\mathrm{tr}(X_\mu X^\mu)]^2\nonumber\\
&&\qquad\qquad+g_1B_{\mu\nu}\mathrm{tr}(\tau^3\overline{W}^{\mu\nu})+2i\mathrm{tr}(X_\mu
X_\nu\overline{W}^{\mu\nu}) +ig_1B_{\mu\nu}\mathrm{tr}(\tau^3X^\mu
X^\nu)\bigg]\;,\\
&&i\mathcal{K}_{14}^{{\rm
1F},\Sigma\neq0}\mathrm{tr}_f(\tilde{A}_\mu \tilde{A}_\nu
V^{\mu\nu})=i\mathcal{K}_{14}^{{\rm 1F},\Sigma\neq0}\bigg[
\frac{1}{2}\mathrm{tr}(X_\mu X_\nu\overline{W}^{\mu\nu})
+\frac{1}{4}g_1B_{\mu\nu}\mathrm{tr}(\tau^3X^\mu
X^\nu)+\frac{i}{4}[\mathrm{tr}(X_\mu
X_\nu)]^2\nonumber\\
&&\qquad\qquad-\frac{i}{4}[\mathrm{tr}(X_\mu X^\mu)]^2 \bigg]\;.
\end{eqnarray}
Comparing the above results with the standard electroweak chiral
Lagrangian \cite{Appelquist:1980vg,Zhang:2007zi}, we find
\begin{eqnarray}
&&f^2=4(F_0^{\rm 1F})^2\;,\qquad \beta_1=0\;,\qquad
\alpha_1=2(\mathcal{K}_{2}^{{\rm
1F},\Sigma\neq0}-\mathcal{K}_{13}^{{\rm 1F},\Sigma\neq0})\;,\qquad
\alpha_2=\alpha_3=-\mathcal{K}_{13}^{{\rm
1F},\Sigma\neq0}+\frac{\mathcal{K}_{14}^{{\rm
1F},\Sigma\neq0}}{4}\;,\nonumber\\
&&\alpha_4=\frac{\mathcal{K}_{4}^{{\rm
1F},\Sigma\neq0}+2\mathcal{K}_{13}^{{\rm
1F},\Sigma\neq0}-\mathcal{K}_{14}^{{\rm
1F},\Sigma\neq0}}{4}\;,\qquad \alpha_5=\frac{\mathcal{K}_{3}^{{\rm
1F},\Sigma\neq0}-\mathcal{K}_{4}^{{\rm
1F},\Sigma\neq0}-4\mathcal{K}_{13}^{{\rm 1F},\Sigma\neq0}
+2\mathcal{K}_{14}^{{\rm
1F},\Sigma\neq0}}{8}\;,\nonumber\\
&&Z_1=1+g_1^2(2\mathcal{K}_{2}^{{\rm
1F},\Sigma\neq0}+\frac{14}{3}\mathcal{K}_{13}^{{\rm
1F},\Sigma\neq0})\;,\qquad Z_2=1+2g_2^2(\mathcal{K}_{2}^{{\rm
1F},\Sigma\neq0}+\mathcal{K}_{13}^{{\rm 1F},\Sigma\neq0})\;.
\end{eqnarray}
By means of the Schwinger-Dyson analysis employed in Ref.
\cite{Yang:2002he,Zhang:2007zi}, we can obtain the numerical results
of $\mathcal{K}_{i}^{{\rm 1F},\Sigma\neq0}$ and further those of
$\alpha_i$ ($i=1,\cdots,5$). In TABLE \ref{table-alpha1fam}, we list
down the numerical calculation results of $\alpha_i$ for different
kinds of technicolor dynamics.
%%%%%%%%%%%%%%%%%%%%%%%%%%%%%%%%%%%%%%%%%%%%%%%%%%%%%%%%%%%%%%%%%%%%%
\begin{table}[h]
\caption{The obtained nonzero values of the $O(p^4)$ coefficients
$\alpha_1,\alpha_2=\alpha_3,\alpha_4,\alpha_5$ for the one family
technicolor model with the conventional strong interaction QCD
theory values given in Ref. \cite{Yang:2002he} for model A and
experimental values for comparison. Here $\Lambda_{\rm TC}$ and
$\Lambda_{\rm QCD}$ are in TeV, and the coefficients $\alpha_i$ are
in units of $10^{-3}$.} \label{table-alpha1fam} \centering
\begin{tabular}{c c|c c c c}\hline\hline
 $N_{\rm TC}$&$\Lambda_{\rm TC}$&$\alpha_1$ &$\alpha_2=\alpha_3$ &$\alpha_4$
 &$\alpha_5$\\ \hline
3&0.635&$-28.7$&$-10.5$&9.00&$-11.5$\\
4&0.555&$-38.1$&$-13.9$&11.9&$-15.2$\\
5&0.499&$-47.2$&$-17.2$&14.8&$-18.9$\\
6&0.458&$-56.8$&$-20.4$&17.7&$-22.6$\\ \hline QCD Theor &
$\Lambda_{\rm QCD}$=0.484$\times10^{-3}$&$-7.06$&$-2.54$&2.20&$-2.81$\\
QCD Expt &&$-6.0\pm0.7$&$-2.7\pm0.4$&$1.7\pm0.7$&$-1.3\pm1.5$\\
 \hline
\end{tabular}
\end{table}
%%%%%%%%%%%%%%%%%%%%%%%%%%%%%%%%%%%%%%%%%%%%%%%%%%%%%%%%%%%%%%%%%%
Comparing with the results of the one doublet technicolor model in
Ref. \cite{Zhang:2007zi}, we observe that the present results of the
one family technicolor model also exhibit the $N_{\rm TC}$ scale-up
relations discussed in that paper, and we further find that the
dynamical structure of all the coefficients' dependence of the
technifermion self energy $\Sigma$ in this model is the same as that
in the one doublet technicolor model, except for some overall
factors and a little difference of the detailed form of $\Sigma$ due
to different technifermion numbers (now the number of technifermion
doublet is 4 and the technifermion number is $N_{\rm TF}=8$, while
in the one doublet technicolor model there is only 1 technifermion
doublet and the technifermion number is 2). The extra small
difference caused by the difference of technifermion numbers can be
listed as follows. In this model the $O(p^2)$ coefficient $f$ is
larger by a factor of $2=\sqrt{4}$ than that in the one doublet
technicolor model. And all the $O(p^4)$ coefficients, except for
$Z_1$ and $Z_2$, are also larger by a factor 4. This extra factor 4
is due to the large $N$ analysis, from which we know $f$ is order of
$\sqrt{N}$ and $\alpha_i$ ($i=1,2,3,4,5$) are order of $N$. Note
that from experiments we must keep fixed $f=250$GeV, which implies
$F_0^{\rm 1F}=125$GeV. As already discussed in Ref.
\cite{Zhang:2007zi}, the reducing of $F_0$ by one half is equivalent
to the reducing of $\Lambda_{\rm TC}$ by one half, which has no
effect on the coefficients $\mathcal{K}_i^{\Sigma\neq 0}$
($i=1,2,3,4,13,14$). Thus the coefficients $\alpha_i$ we obtained in
this model are roughly four times as large as those in the one
doublet technicolor model, if we ignore the small differences caused
by the different $\Sigma$ due to different technifermion numbers.
The above numerical results agree well with the $N_{\rm TC}$ and
$N_{\rm TF}$ scale-up assumption of the coefficients of the EWCL
argued by Ref. \cite{Hill:2002ap}. This strongly confirms the
reliability of our dynamical computation formalism. Moreover, we
have systematically obtained all the explicit values of the $O(p^4)$
coefficients of the EWCL from this one family technicolor model.
These results are hitherto absent in the literature so far as we
know. The dynamical computation formalism used in this paper may be
employed in many other underlying models beyond the SM, and this
kind of work is still in preparation.

\begin{acknowledgments}
This work is supported by the National Natural Science Foundation of
China under Grant No.10747165, and Sun Yet-Sen University Science
Foundation.
\end{acknowledgments}

%%%%%%%%%%%%%%%%%%%%%%%%%%%%%%%%5

\end{document}